\documentclass{article}
\usepackage{spconf,amsmath,graphicx}
\usepackage{comment}
\usepackage{epstopdf}           

\usepackage{comment}
\usepackage{verbatim}
\usepackage{soul, xcolor}
\usepackage{amsfonts}
\usepackage{booktabs}
\usepackage{multirow}
\usepackage[noadjust]{cite}

\title{\vspace{-2mm} Deep neural network Based Low-latency Speech Separation with Asymmetric analysis-Synthesis Window Pair \vspace{-2mm}}
\name{Shanshan Wang\textsuperscript{*}\thanks{The authors wish to thank CSC-IT Centre of Science Ltd., Finland,  for providing computational resources  and Paul Magron for helping with illustrations included in this paper. \newline *  authors contributed equally}, Gaurav Naithani\textsuperscript{*}, Archontis Politis, Tuomas Virtanen \vspace{-5mm}}
\address{Audio Research Group, Tampere University, Tampere, Finland \\ \vspace{-2mm}
 \normalsize Email:\{ shanshan.wang, gaurav.naithani, archontis.politis, tuomas.virtanen\}@tuni.fi
}
\begin{document}
%
\maketitle
\vspace{-5mm}
\begin{abstract}
Time-frequency masking or spectrum prediction computed via short symmetric windows are commonly used in low-latency deep neural network (DNN) based source separation. In this paper, we propose  the usage of an asymmetric analysis-synthesis window pair which allows for training with targets with better frequency resolution, while retaining the low-latency  during inference suitable for real-time speech enhancement or assisted hearing applications. In order to assess our approach across various model types and datasets, we evaluate it with both speaker-independent deep clustering (DC) model and a speaker-dependent mask inference (MI) model.  We report an improvement in separation performance of up to 1.5 dB in terms of source-to-distortion ratio (SDR) while maintaining an algorithmic latency of 8 ms.

\end{abstract}
\vspace{0mm}
\section{Introduction}
\label{sec:intro}
DNN-based methods have now become the current state-of-the-art in various signal processing problems including monaural speech separation \cite{wang2018supervised,huang2014deep, erdogan2015phase, yu2017permutation,hershey2016deep, Isik+2016, wang2018alternative, luo2019conv, liu2018casa}. These can be broadly divided into two categories: time-frequency (TF) spectrum based techniques \cite{huang2014deep, erdogan2015phase, yu2017permutation, hershey2016deep, liu2018casa}, which map the TF representation (e.g., short-time Fourier transform (STFT)) of an input acoustic mixture to the output TF representation of the constituent clean speech,  and, end to end learning based techniques, which directly map a mixture waveform to separated speech waveforms \cite{venkataramani2018end, luo2019conv, tasnet2018}. In the case of the former, the algorithmic latency of the system is restricted by the choice of synthesis window used for overlap-add reconstruction of the output. Applications like hearing-aids \cite{bramslow2010preferred} and cochlear implants \cite{hidalgo2012low} are very restrictive in terms of allowable latencies. Especially in hearing aids, the presence of two paths through which the user receives the sound,  the direct path and the path through the hearing aid, leads to the user experiencing disturbances \cite{stone2008tolerable, agnew2000hearing}.

In general, TF spectrum based DNN speech separation methods have been using the same symmetric analysis and synthesis windows.  The algorithmic delay of these methods is limited by length of the synthesis window. For low-latency applications (e.g., 5-10 ms), window lengths used in conventional speech processing (e.g., 20-40 ms   \cite{paliwal}) cannot be used. Using a very short window implies poor frequency resolution in the TF representation, resulting in loss of fine spectral structure during stationary speech segments. This loss, in turn, makes the task of the DNN to separate speakers harder as  disjoint spectral detail of different speakers at a fine resolution is smeared and overlapped at a lower resolution.

In this paper, we postulate that the use of the same analysis-synthesis short window is a sub-optimal choice in DNN-based speech separation systems catering to low-latency applications. We instead utilize an asymmetric windowing scheme first proposed in \cite{mauler2007low} where a larger analysis window is used, similar to the ones used in conventional processing,  yielding a good frequency resolution. The synthesis window is however shorter allowing low latency operation. The windowing scheme offers perfect reconstruction when no intermediate processing is involved.

 In order to show the independence of our proposed approach from the type of models and datasets used,  we evaluate it on two tasks: speaker-independent separation with an online DC model \cite{wang2019low}, and speaker-dependent separation with mask inference (MI) network that directly predicts masks. We evaluate the former and latter on two-speaker mixtures from Wall Street Journal (WSJ0) \cite{garofalo2007csr} and Danish HINT \cite{nielsen2011danish, nielsen2009development} databases, respectively. Improvement in separation performance measured by SDR \cite{vincent2006performance}, source-to-interference ratio (SIR), and source-to-artifact ratio (SAR) is observed.
 We report an improvement of up to 1.5 dB in terms of the SDR for models with an asymmetric analysis-synthesis window pair over the baseline models with a symmetric windowing scheme. The short time objective intelligibility (STOI) and perceptual evaluation of speech quality (PESQ) scores \cite{pesq} are reported as well.

\begin{figure}[t!]
\centering
\includegraphics[scale=0.35]{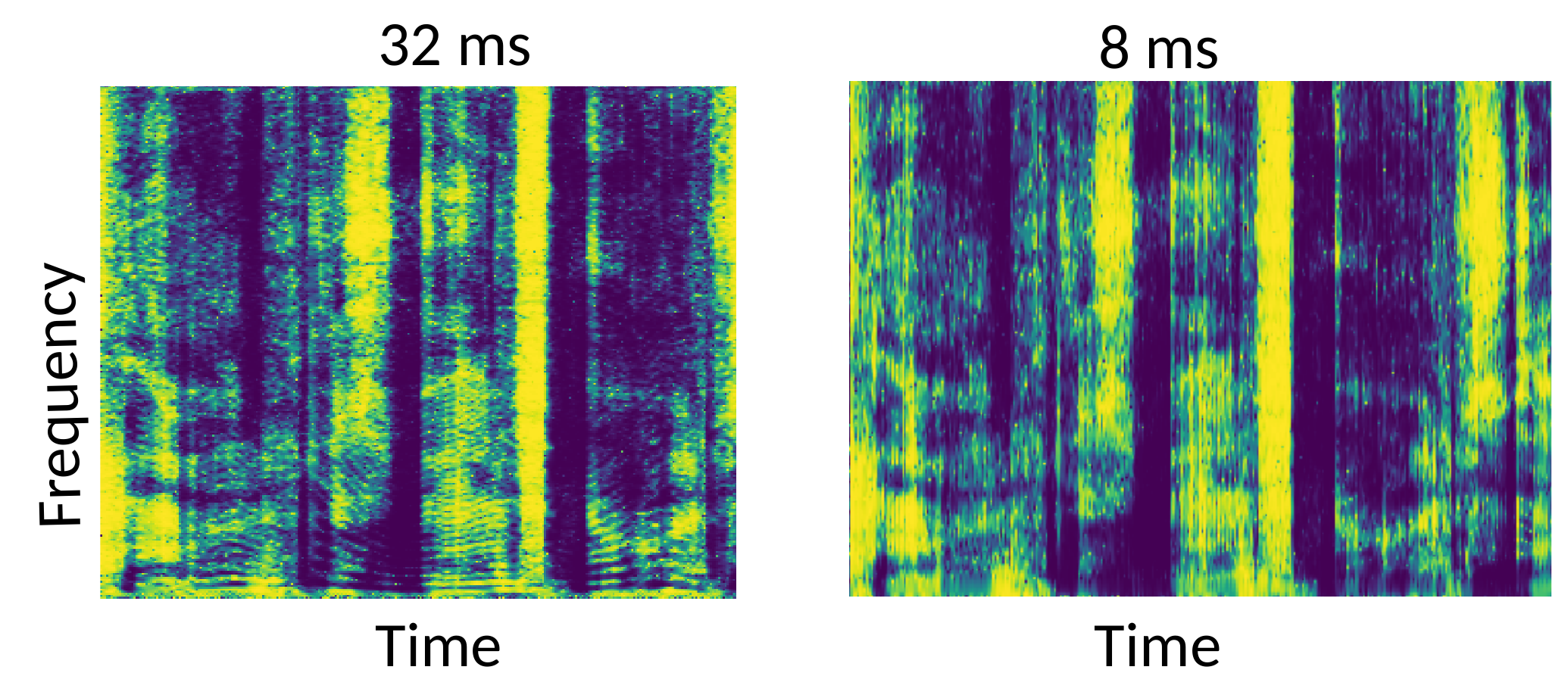}
  \caption{\small An example of oracle masks calculated with symmetric Hann windows of length 32ms (left) and 8 ms (right).}
  \label{fig:masks}
\end{figure}

\begin{figure}[b!]
\centering
\includegraphics[scale=0.28]{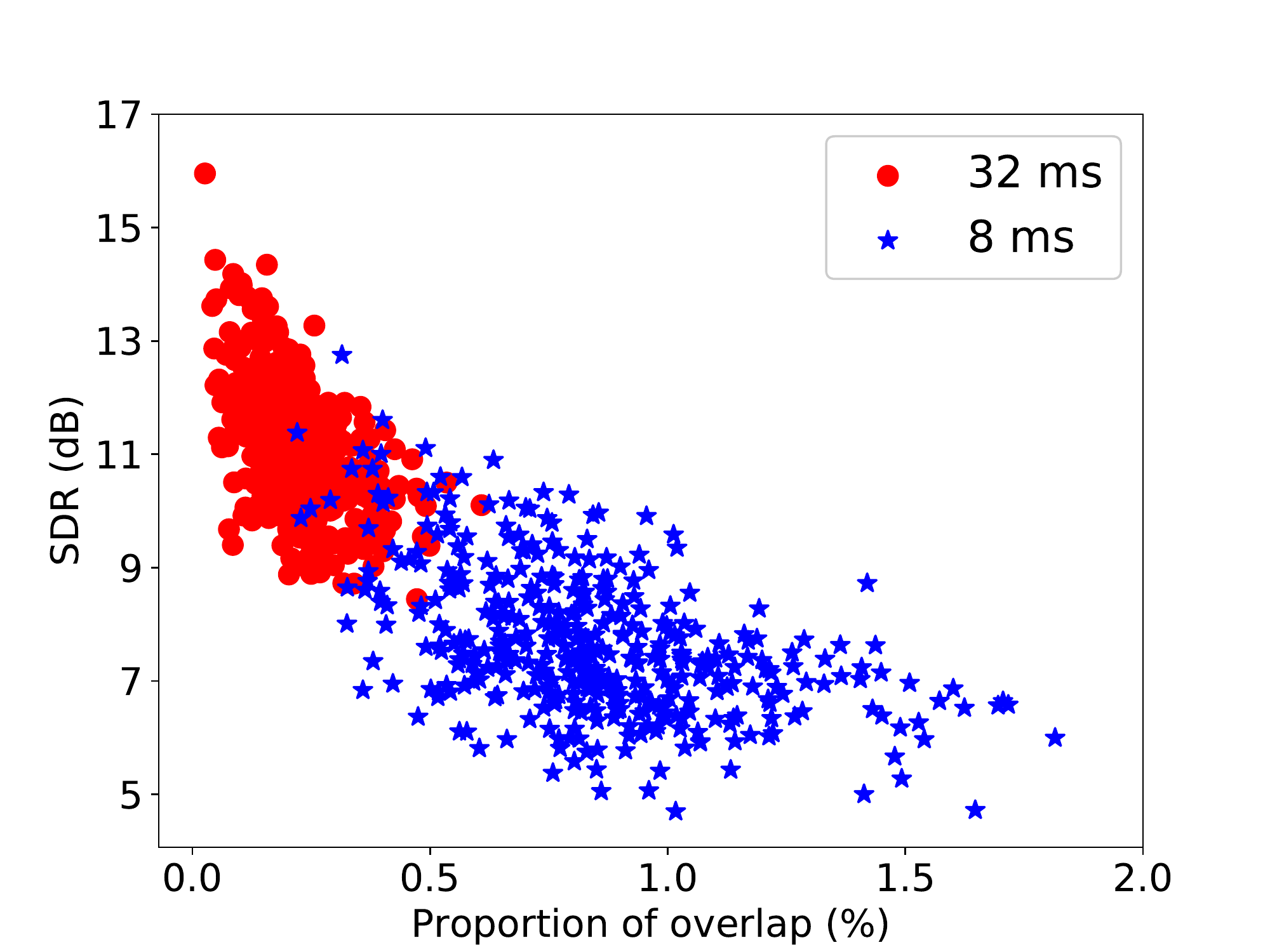}
  \caption{ \small Oracle SDR (dB) and proportion of overlap (\%) of sources for symmetric Hann windows of length 32 ms and 8 ms.}
  \label{fig:overlap}
\end{figure}

\begin{figure}[t!]
\centering
\includegraphics[scale=0.52]{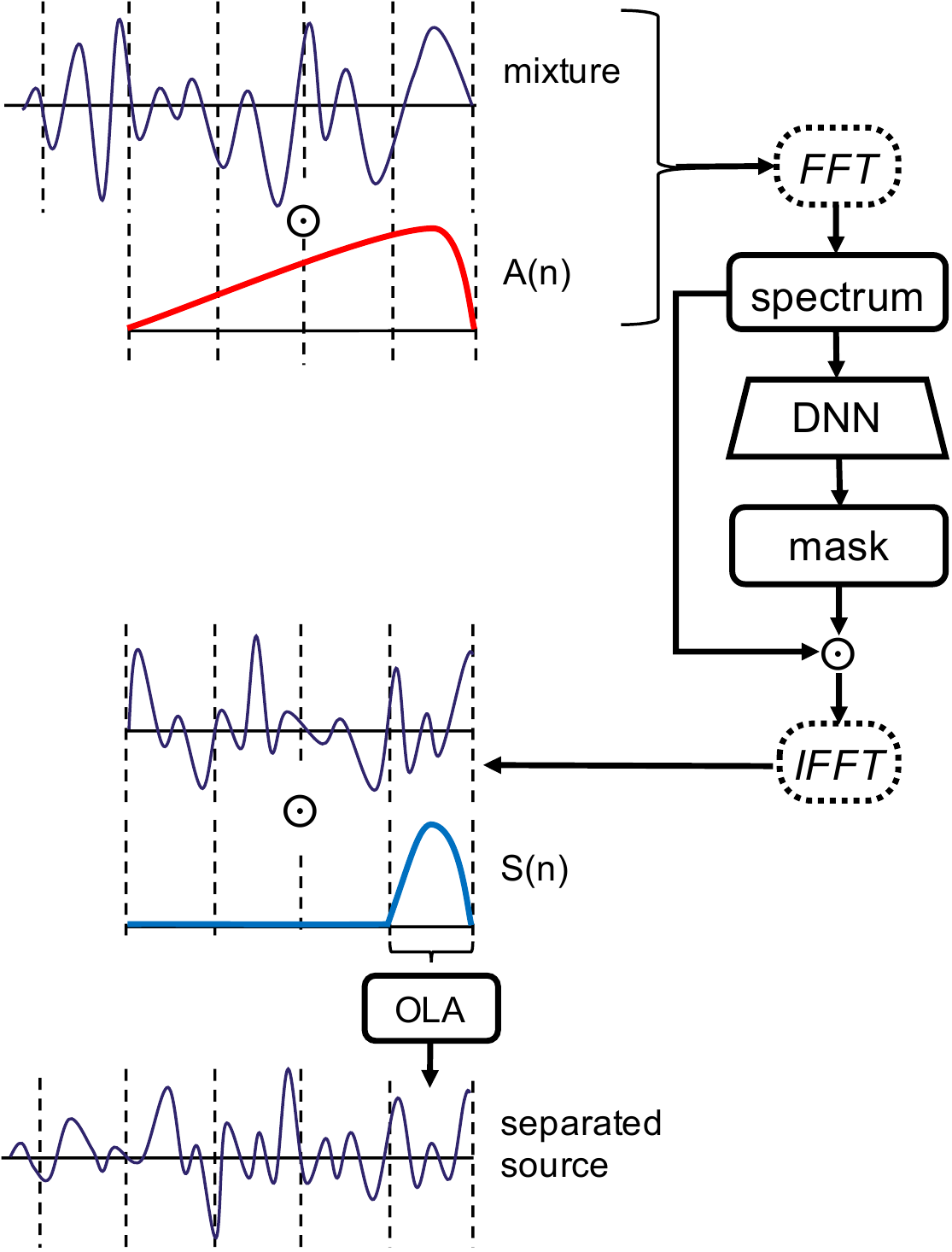}
  \caption{\small Illustration of an asymmetric analysis/synthesis window pair in a general DNN-based speech separation (MI) system. A single $K$ frame length  mixture is used to estimate $2M$ length separated speech. \small$\odot$  \hspace{0.5mm} denotes element-wise multiplication.}
  \label{fig:diagram_sd}
\end{figure}

\vspace{-2mm}

\section{Proposed method }\label{sec:method}
In TF spectrum based speech separation with DNNs, the training targets  for supervised learning are usually in the form of a TF representation, e.g.,  TF masks or affinity matrices (for deep clustering \cite{hershey2016deep}, \cite{wang2019low}). The STFT is a popular choice where the choice of the window length is important. There is a trade-off between better frequency resolution with longer windows, and better temporal resolution with shorter windows. Fig. \ref{fig:masks} shows  two ideal ratio masks computed with window lengths of 32 and 8 ms. It can be seen that masks with 32 ms retain the harmonic structure of the speaker at low frequencies, while the same structure is smeared with the 8 ms window.  This also implies there is more spectral overlap between the sources and hence the targets chosen for DNNs are more challenging to learn.
Fig. \ref{fig:overlap} shows a scatter plot  of oracle SDR values of mixtures  from the Danish HINT dataset \cite{nielsen2011danish, nielsen2009development} and proportion of overlap for window lengths of 32 and 8 ms. The proportion of overlap is given by $\frac{\tilde{N}}{N},$ where $\tilde{N}$ and $N$ are number of TF bins where both sources are active and total number of TF bins, respectively. $\tilde{N}$ is calculated as
$\sum (S_1 \geq \tau * s^{max}_{mix}) \land (S_2 \geq \tau * s^{max}_{mix})$, where $S_1$, $S_2$ and $s^{max}_{mix}$ are  source 1, source 2 magnitude STFT, and maximum of magnitude mixture STFT. $\land$ is logical \textit{and} operator  and $\tau$ is chosen to be 0.1. As Fig. \ref{fig:overlap} shows, the larger the proportion of overlap,  the lesser the SDR is observed.

\subsection{Low-latency separation using asymmetric windows}
In TF-based speech separation, a TF representation, e.g. STFT, is first computed from chunks of input mixture windowed using an analysis window. The STFT features are then fed to a DNN to get a TF mask or spectrum corresponding to the constituent speakers, either directly or via some clustering step as in \cite{hershey2016deep}.  The estimated spectrum is then converted to time domain using the inverse fast Fourier transform (IFFT), multiplied by a synthesis window, and then overlap-added with previous output frames. As the latency is determined by the length of the synthesis window, the poor frequency resolution can be mitigated by using a larger analysis and shorter synthesis  window provided they fulfill the Princen-Bradley conditions \cite{BosiGoldbergIntroToDACS2003} for perfect reconstruction.  Such an asymmetric windowing design is presented in \cite{mauler2007low} in the context of adaptive filtering for speech enhancement. In this section, firstly, we discuss how asymmetric windowing is applied in a low-latency separation system and later discuss the  asymmetric windowing scheme.

The diagram of the whole process is depicted in Fig.~\ref{fig:diagram_sd}. The mixture and its constituent sources, $\textit{source}_1$ and $\textit{source}_2$,  are first divided into $K$-sample frames with $M$-sample overlap. Each of these $K$-sample frames is then multiplied by an  analysis window of the same length. The single-frame spectral features are calculated by the fast Fourier transform (FFT). The  FFT magnitude features of the mixture are then fed into DNN to predict the masks corresponding to the constituent speakers, either directly or implicitly. The latter is the case with DC model where embeddings corresponding to the TF bins are first outputted, which are then converted into masks after a clustering step. It should be noted that the target masks computed here correspond to  features computed using a $K$-sample analysis window.
The mixture features are multiplied by the corresponding masks estimated by the DNN to give the estimated speech spectrum which are converted back to the time domain speech via the IFFT. Finally, a   synthesis window  of length $2M$ zero-padded to length $K$, is  applied to the output time-domain frame before overlap-add, to get the separated source signal.

%



\subsection{Asymmetric windowing}
 Mauler and Martin \cite{mauler2007low} reported asymmetric windowing schemes intended for real-time low-latency speech enhancement. It allowed for adequate frequency resolution during estimation of speech statistics, while keeping a relatively short synthesis window. The same principle is adopted in this work for DNN-based source separation. For an STFT with a hop size $M$, the windowing scheme in \cite{mauler2007low} is based on a Hann window prototype $H_{2M}(n)$ of length $2M$, defined as,
\begin{equation}
   H_{2M}(n) = 0.5(1-\text{cos}(\pi\frac{n}{M})),\quad n=0,...,2M-1 \enskip \text{.}
\end{equation}

\begin{figure}[t!]
\centering
\includegraphics[scale=0.6]{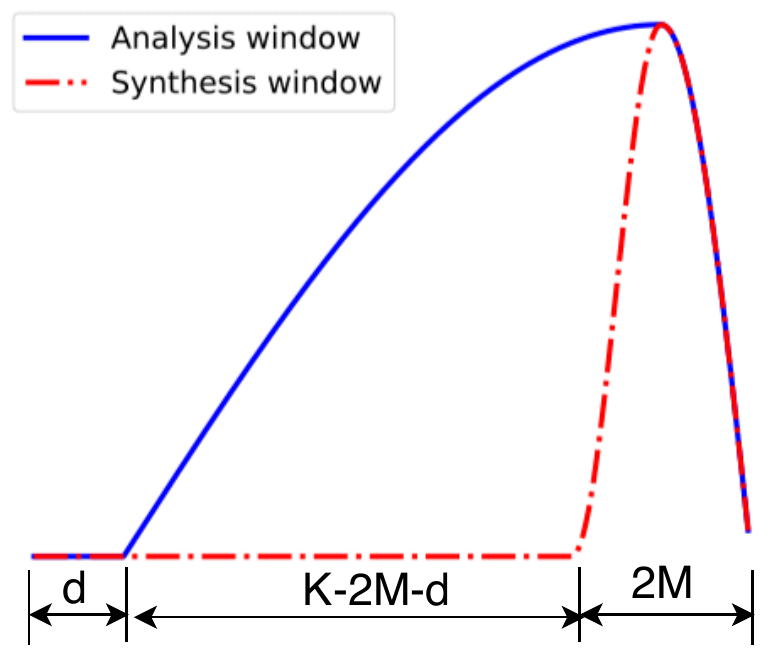}
  \caption{\small The asymmetric analysis and synthesis windows. $K$ and $2M$ denote the lengths of analysis and synthesis window, respectively. The synthesis window is zero-padded to length $K$. }
  \label{fig:asy_win}
\end{figure}

 By defining an analysis-synthesis window pair \{$A(n), S(n)$\} of lengths $K$ and $2M$, respectively, with $K>2M$, both windows have their last length-$M$ segment as the root of the Hann prototype given by,

\begin{equation}
\small
\resizebox{0.9\columnwidth}{!}{
$
A(n)=S(n)=\sqrt{H_{2M}(n-K+2M)},\; K-M\leq n < K  \, \text{.}
$
}
\end{equation}
The first asymmetric segment of the analysis window is generated from another longer half Hann window prototype as,

\begin{align}
\small
\label{eq:h(n)}
\resizebox{0.9\columnwidth}{!}{
$A(n)  = \begin{cases}
0 \, , &  $0$ \leq  n  < d   \\
\sqrt{H_{2(K-M-d)}(n-d)}\, , &   d\leq n < K-M   \text{,}
\end{cases}$
}
\end{align}
\noindent where the first $d$ samples are zeros to mitigate aliasing effects (please refer to \cite{mauler2007low} for more details). Finally, imposing a perfect reconstruction constraint on the pair, in which their product should result in the original Hann prototype $H_{2M}(n)$,
 \begin{equation}
 \small
A(n)S(n)=
 \begin{cases}
0 \, ,         &  $0$ \leq  n  < K-2M\\
H_{2M}(n-K+2M) ,  &  K-2M \leq n < K,
\end{cases}
\end{equation}
we obtain the first segment of the shorter synthesis window as,
 \begin{align}
 \small
\label{eq:h(n)}
S(n)  = \begin{cases}
0  \, ,                         &   $0$\leq n<K-2M \\
\frac{H_{2M}(n-K+2M)}{A(n)} \, ,    &  K-2M\leq n<K-M \text{.}
\end{cases}
\end{align}
The various window segments for the window pair are depicted in Fig.~\ref{fig:asy_win}. 

\section{EVALUATION} \label{sec:evaluation}
In order to show the generality of the proposed method, we evaluate it in two main parts: \textit{a)} speaker-independent separation with DC \cite{wang2019low} on  WSJ0, and, \textit{b)} speaker-dependent separation with direct mask inference (MI) on Danish HINT~\cite{naithani2018deep}. For the former, we consider offline and online separation separately depending upon the length of audio available for cluster estimation corresponding to the constituent speakers. Offline separation implies that the entire signal is available and online separation implies that a certain length in the beginning of the signal, referred to as \textit{buffer length} (0.6 s in this work), is used for estimating fixed cluster centres which are then used to cluster the embeddings for the rest of the signal~\cite{wang2019low}.

\begin{figure}[b!]
\centering
\includegraphics[scale=0.30]{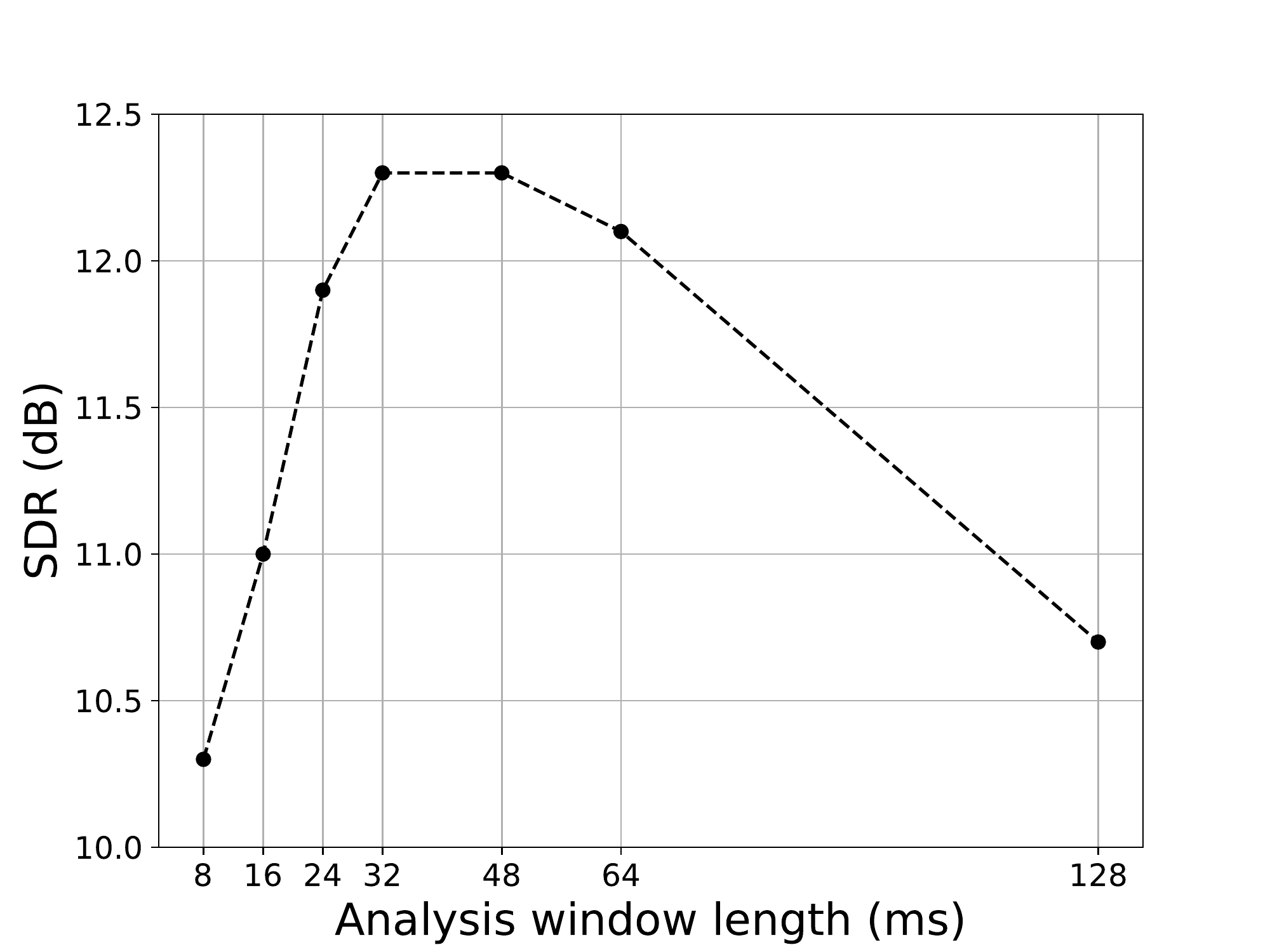}
  \caption{\small The separation performance, measured in terms of SDR, for different analysis window lengths for oracle IBM from WSJ0.}
  \label{fig:sdr_vs_win}
\end{figure}

\subsection{Experiments}

For~(\textit{a}),  a four-layer long-short-term-memory (LSTM) with 600 units in each layer is used, followed by a feedforward layer with \textit{tanh} activation~\cite{wang2019low}.  The network predicts 40-dimensional embeddings for each of the TF bins, which are then clustered using K-means to get the ideal binary masks corresponding to the constituent speakers. The embedding vectors are normalized to unit norm. The cost function being optimized here is a low rank formulation of squared $L2$ loss between the ideal and estimated binary affinity matrices. For~(\textit{b}) a three-layer LSTM network with 512 units in each layer is used, followed by a feedforward layer with \textit{sigmoid} activation  as in \cite{naithani2018deep}. The cost function being minimized here is $L2$ loss between estimated and ideal ratio masks.
In order to find the best asymmetric analysis window length, different resolutions are investigated, similar to  \cite{vincent2007oracle}. Fig.~\ref{fig:sdr_vs_win} depicts oracle performance in terms of SDR for 300 mixtures from the WSJ0 dataset for different asymmetric analysis window lengths. It should be noted that the synthesis window length is fixed at 8 ms in all cases. The best SDR is achieved for 32 ms and 46 ms asymmetric analysis window length. Taking the computational complexity into account, 32 ms window is chosen for the rest of the experiments. The PyTorch \cite{paszke2017automatic} framework is used to train the networks and the Adam optimizer \cite{kingma2014adam} with default parameters is used. Early stopping  \cite{giles2001overfitting} of the training is done if no improvement in validation loss is observed for 15 consecutive epochs.


\begin{table}[t!]
\centering
\caption {The oracle/evaluation metrics (dB) for DC with different types of windows and window lengths (ms). Sym. and Asym. denote symmetric and asymmetric windows, respectively. ($A$, $S$) denotes analysis and synthesis window lengths. }
\label{tab:eval_DC}
\vspace{2ex}
\small
\vfill
\begin{tabular}{l@{\hspace{2ex}}c@{\hspace{1ex}}c@{\hspace{2ex}}c@{\hspace{2ex}}c@{\hspace{2ex}}c@{\hspace{2ex}}c@{\hspace{2ex}}r}
\toprule
Mode                     & Window   & ($A$, $S$)    & SDR & SIR & SAR & STOI & PESQ \\ \midrule
\multirow{3}{*}{Oracle}  & Sym.      &  (32, 32)    & 13.7    & 22.8    & 14.4 & 0.94 & 3.31     \\
                         & Sym.      &  (8, 8)      & 10.3    & 19.5    & 11.0 & 0.91 & 2.78   \\
                         & Asym.     &  (32, 8)     & \textbf{12.3}    & \textbf{20.5}    & \textbf{13.2} & \textbf{0.94} & \textbf{3.15}  \\  \midrule

\multirow{3}{*}{Offline} & Sym.      &  (32, 32)    & 8.0    & 16.2    & 9.3 & 0.83 & 1.90     \\
                         & Sym.      &  (8, 8)      & 6.6    & 14.9    & 7.9 & 0.81 & 1.86     \\
                         & Asym.     &  (32, 8)     & \textbf{7.4}    & \textbf{15.1}    & \textbf{8.8}
                         & \textbf{0.82} & \textbf{1.86}
                         \\  \midrule
\multirow{2}{*}{Online}  & Sym.      &  (8, 8)      &  5.7   & 13.5    & 7.3 & 0.80 & 1.82      \\
                         & Asym.      &  (32, 8)     &  \textbf{7.1}   & \textbf{14.5} & \textbf{8.6} & \textbf{0.82} & \textbf{1.85}
\\ \bottomrule
\end{tabular}
\end{table}

\begin{table}[b!]
\centering
\caption { The evaluation metrics (dB) for MI model   with different types of windows and window lengths (ms).}
\label{tab:danish_results}
\vspace{2ex}
\small
\vfill
\begin{tabular}{l@{\hspace{2ex}}c@{\hspace{2ex}}c@{\hspace{2ex}}c@{\hspace{2ex}}c@{\hspace{2ex}}c@{\hspace{2ex}}c@{\hspace{2ex}}r}
\toprule
Mode & Window  & ($A$, $S$) & SDR & SIR & SAR & STOI & PESQ\\
 \midrule
 \multirow{3}{*}{Oracle} & Sym. & (32, 32) & 11.6 & 15.1 & 14.4 & 0.97 & 3.55 \\
                         & Sym. & (8, 8) & 8.0 & 11.0 & 11.6 & 0.94 & 2.63 \\
                         & Asym.& (32, 8) &  \textbf{10.1} & \textbf{13.0} & \textbf{13.5} & \textbf{0.95} & \textbf{2.99} \\ \midrule
 \multirow{3}{*}{Online} & Sym. & (32, 32) & 10.2 & 13.6 & 13.2 &0.91 & 2.68 \\
                         & Sym. & (8, 8) & 7.3 & 10.0 & 11.2  & 0.89 & 2.33\\
                         & Asym. & (32, 8) &  \textbf{8.8} & \textbf{11.6} & \textbf{12.5} & 0.90 & \textbf{2.48}\\
\bottomrule
\end{tabular}
\end{table}

\subsection{Dataset}
Two-speaker synthetic mixtures  are created from WSJ0/Danish HINT for speaker-independent and speaker-dependent separation, respectively. WSJ0 has 101 and 18 speakers for training/validation and testing data, respectively.
The training, validation and  testing data consists of 20000 ($\sim$30 hrs), 5000 ($\sim$8 hrs), and 3000 ($\sim$5 hrs) mixtures, respectively.  The speakers in the test data  are different from the ones in the training/validation set. The mixtures are formed by first removing silence in the beginning of the constituent signals and then summing, to ensure that both speakers are active during the \emph{buffer length} similar to~\cite{wang2019low}.  All mixtures are downsampled to 8 kHz for reducing computational burden. For Danish HINT,  we choose three speaker pairs, M1M2, F1F2, and M1F1. Each speaker has 13 lists each consisting of 20 five-word sentences of natural speech. Eight lists for training (L6-L13),  two lists for validation (L4, L5) and two lists for testing (L1, L2) are used as was done in~\cite{naithani2018deep}. The mixtures are downsampled to 16 kHz.

\subsection{Results and discussion}
The  baseline corresponds to processing with a conventional low-latency symmetric analysis/synthesis window of 8 ms. 32 ms symmetric analysis/synthesis pair serves as the ceiling on separation performance. Table \ref{tab:eval_DC} shows the separation metrics for  speaker-independent separation with DC along with the oracle performance. Our approach is shown in bold. The oracle performance improves by 2 dB in terms of SDR for asymmetric windowing scheme compared to the baseline. For offline DC and online DC, an improvement of 0.8 dB and 1.4 dB in SDR, respectively,  is observed. It is notable that with asymmetric windowing,  online DC not only performs better than the online DC baseline but also outperforms offline DC baseline. Table  \ref{tab:danish_results} shows the separation metrics for the Danish HINT dataset along with the oracle performance. The metrics corresponding to the three speaker pairs have been averaged. An improvement of about 1.5 dB in terms of SDR over the baseline is observed. The improvements in STOI and PESQ scores are reported as well.

 Moreover,  we verify  that these improvements are due to a better ground truth for DNN training rather than a better input representation in the form  of a better resolution input. For this, we train the network with target masks computed with a resolution corresponding to the synthesis window. The results are shown in Table~ \ref{tab:danish_results_v2},  $\mathbb{M}_8$  and  $\mathbb{M}_{32}$ are ratio masks  computed corresponding to resolution 8 ms and 32 ms, respectively. The 8 ms synthesis window is used during inference for all cases. It can be seen that separation performance with  8 ms asymmetric synthesis window target is similar to that with 8 ms symmetric synthesis window target, which confirms our hypothesis.

\begin{table}[t!]
\centering
\caption { The evaluation metrics (dB) for Danish HINT  with different input/target resolutions while training MI model.}
\label{tab:danish_results_v2}
\vspace{2ex}
\small
\vfill
\begin{tabular}{l@{\hspace{1ex}}c@{\hspace{1ex}}c@{\hspace{1ex}}c@{\hspace{1ex}}c@{\hspace{1ex}}c@{\hspace{1ex}}c@{\hspace{1ex}}r}
\toprule
 & Input  & Target & SDR & SIR & SAR & STOI & PESQ\\
 \midrule
 \multirow{1}{*}{$\mathbb{M}_8$}  & (Sym., 8)& (Sym., 8) &  7.3 & 10.0 & 11.2 & 0.89 & 2.33  \\
 \multirow{1}{*}{$\mathbb{M}_{32}$}  & (Asym., 32) & (Asym., 32) &  \textbf{8.8} & \textbf{11.6} & \textbf{12.5} & \textbf{0.90} & \textbf{2.48} \\
\bottomrule
\end{tabular}
\end{table}

\section{CONCLUSION} \label{sec:conclusion}
In this  paper, we  propose to use asymmetric analysis/synthesis pairs for low-latency DNN-based speech separation. We evaluate it for a speaker-independent DC model and speaker-dependent MI model.  We report an improvement of up to 1.5 dB in terms of SDR in our evaluation. We also note that the improvement is independent of the types of models/datasets used. In addition, we confirm that improvement in performance is on account of better ground truths to train DNNs with the proposed asymmetric windowing scheme.






\bibliographystyle{IEEEtran}
\small
\bibliography{refs}

\end{document}